\documentclass[a4, 10point]{article}
\usepackage[top=2cm, bottom=2cm, left=2.5cm, right=2.5cm]{geometry}

\usepackage[latin1]{inputenc}
\usepackage{amsfonts}
\usepackage{amsmath}
\usepackage{amssymb}
\usepackage{amsthm}
\usepackage[T1]{fontenc}
\usepackage{graphicx}

\theoremstyle{definition} 
\theoremstyle{definition} \newtheorem{defin}{Definition}
\theoremstyle{plain} \newtheorem{thm}{Theorem}[section]
\theoremstyle{plain} \newtheorem{lem}{Lemma}[section]
\theoremstyle{plain} 
\theoremstyle{plain} \newtheorem{prop}{Property}[section]
\theoremstyle{plain} 

\title{\sc{On graph parameters guaranteeing fast Sandpile diffusion}}
\author{Ayush Choure \quad Sundar Vishwanathan \\ Department of Computer Science and Engineering  \\ Indian Institute of Technology, Bombay \\ ayush, sundar@cse.iitb.ac.in}

\begin{document}
\maketitle

\begin{abstract}

The Abelian Sandpile Model is a discrete diffusion process defined on graphs (Dhar \cite{DD90}, Dhar et al. \cite{DD95}) which serves as the standard model of \textit{self-organized criticality}. The transience class of a sandpile is defined as the maximum number of particles that can be added without making the system recurrent (\cite{BT05}). We demonstrate a class of sandpile which have polynomially bound transience classes by identifying key graph properties that play a role in the rapid diffusion process. These are the volume growth parameters, boundary regularity type properties and non-empty interior type constraints. This generalizes a previous result by Babai and Gorodezky (SODA 2007,\cite{LB07}), in which they establish polynomial bounds on $n \times n$ grid. Indeed the properties we show are based on ideas extracted from their proof as well as the continuous analogs in complex analysis.  We conclude with a discussion on the notion of degeneracy and dimensions in graphs.


\end{abstract} 

\section{Introduction}

The abelian sandpile model(ASM) is a type of discrete diffusion process defined on graphs. The model was pioneered by Dhar \cite{DD90} as means of investigating the phenomena known as \textit{self-organized criticality}, in the dynamics of sandpile formation. A close cousin is the celebrated loop-erased random walks model \cite{LL}. If one plots the length of a loop-erased random walk, on the same underlying graph, against time, the results obtained are qualitatively similar to the one obtained from observing sandpile weights against time. Indeed the similarity extends into numerous seemingly different phenomena like stress distribution in earthquakes, size distribution in raindrops, etc (see the recent comprehensive survey article by Dhar \cite{DD06}).


In the standard sandpile model, ``sand particles'' are added at vertices of a (multi)graph. A site (vertex) is stable as long as the number of particles at the site are less than its degree. Adding any more particles makes the site unstable and is accompanied by the unstable site passing a particle along each incident edge to its neighboring sites. This relaxation process is called {\em{toppling}}. There is a special site called the \textit{sink} which cannot topple. To ensure that every relaxation process eventually stabilizes, one needs the condition that the sink is reachable from every other site. In the course of evolution via particle additions and toppling, the sandpile goes through a sequence of configurations. Those which can be revisited in any toppling sequence are called {\em{recurrent}}, the remaining ones are termed {\em{transient}}. Typically, one starts with the empty configuration and as particles are added, one moves through transient configurations till a recurrent configuration is reached. Thereafter the configurations stay recurrent. The steady state behavior of a sandpile is characterized by its set of recurrent states. It has been observed by physicists that for most natural phenomena, the time taken to reach a recurrent state is small. Hence any acceptable model must reflect this behavior and an important parameter of study for these systems is the time taken to reach recurrence.

The parameter of importance in our discussion will be the number of particles needed to reach recurrence, when particles are added adversatively. The strategy of adding particles at uniformly random sites yields easily to a simple coupon collector type argument, thereby resulting in polynomial bounds on the \textit{expected} time to recurrence (as previously noted in \cite{BT05}). However, in the case of adversarial site selections, the problem acquires a distinctly potential theoretic flavor. In this scenario, our goal is to add particles so as to avoid a recurrent state for as long as possible. This problem was highlighted by Babai and Toumpakari \cite{BT05} where they define this number of particles as the \textit{transience class} of the sandpile. This later motivated the insightful work by Babai and Gorodezky \cite{LB07} on grid based sandpiles. In this breakthrough paper, Babai et. al. show that for the standard $n \times n$ square grid based sandpile, the maximum number of particles one can add before hitting a recurrent state is $O(n^{30})$, later improved to $O(n^{15})$ \cite{LBpc}. Using arguments based on LP-duality, harmonic functions, and symmetry, Choure and Vishwanathan \cite{CS11}\cite{CS12} improve this bound to $O(n^7)$. They also establish a sandwich theorem for the transience class of any sandpile in terms of the values of harmonic functions, showing a tight correspondence between the random walks properties and sandpile diffusions.

The relevance of grid sandpile comes from the fact that they serve as the standard discrete substitutes of planar regions in both statistical physics and traditional engineering research in finite element methods. It is therefore natural to ask if other graphs which can arise as regular (or irregular even) tessellations of plane (or higher dimensional spaces) also obey polynomial bounds on their transience classes. More generally, one might not even want to pose restrictions on the integrality of dimension of the space from which these graphs arise. Indeed graphs associated with fractals are quite well analyzed in traditional potential theory. We refer the reader to the excellent monograph by Andr\`{a}s Telcs \cite{AT} for a succinct yet thorough introduction to literature in this area. The techniques of Babai and Gorodezky \cite{LB07} extend easily to the honeycomb lattices and triangular lattices. They can also easily be generalized to higher dimension lattices. The problem arises when one wants to deal with graphs without such highly symmetric structure. Our study is aimed at trying to understand the properties which characterize polynomial transience.

\textbf{Our contribution:} We show that if the graph has uniform \textit{polynomial volume growth}, satisfies \textit{mean value property}, and has \textit{high local conductance} of particle percolation, then it satisfies a fundamental superposition property, which is the discrete analogue of the celebrated superposition principal. This property allows one to place particles at a suitable set of multiple nodes and observe the same potential response at a particular site, as one would have by placing all these particles at one site. Using this we prove one of the main results in this paper, the single step Epicenter Propagation Lemma. This is the generalization of the traveling diamond lemma proved in Babai and Gorodezky \cite{LB07}. This lemma demonstrates a very general gluing property of sandpile diffusion. No continuous analogue is known for harmonic functions over continuous regions, where to bound the growth rate of harmonic functions over any path, the best one can do is simply multiply the bounds over each of the balls forming a cover of the path. Using the single step version, we derive the general Epicenter Propagation theorem which effectively demonstrates polynomial bounds on the transience class of large class of sandpile graphs. This result forms the graph theoretic analogue of the classical Harnack's inequality and enables one to prove general bounds on harmonic functions without resorting to limiting arguments linking random walks on graphs to Brownian motions on Euclidean spaces.

We conclude with a discussion on two fundamental questions related to the dimension theory of graphs. The first concerns a dimension like quality of graph which is consistent with the more usual notion of dimension for Euclidean spaces. The second concerns determining if a subgraph is degenerate with respect to the ambient super-graph. In the last section, we discuss some open questions in this area.




\begin{subsection}{Related Work}
\noindent\textbf{Random walks and Sandpile:} The connections between sandpile and random walks have been discussed under various contexts. Dhar, for example, summarizes the connection between the Loop Erased Random Walks and sandpile in \cite{DD06}. We refer the reader to the excellent text by Lawler and Limic \cite{LL} for an introduction to basic properties of LERW. The connection with simple random walks is intuitive and reveals an analogy between properties of sandpile and electric networks. The recent results by Choure and Vishwanathan \cite{CS11} discuss a reduction of the transience class computation problem to that of estimating potentials in a related circuit. These connections are not surprising as the combinatorial Laplacian of the graph of underlying sandpile acts as kind of \textit{generator} for both the models. For an introduction to these potential functions, see Bollob\'{a}s \cite{Bol}. For a discussion on estimating these potentials on graphs coming from geometric settings, the monograph by Telcs \cite{AT} is highly recommended.  For highly related work on potential theoretic approach to analyzing the rotor router model, divisible sandpile and related models, the reader is referred to \cite{LP09}, and \cite{LP10}, for a discussion by Levine and Peres on the spherical asymptotics of diffusions unbounded lattice graphs. The typical question they answer is that when diffusion starts from some point, then the distance from center of the farthest, and the nearest boundary point of the flooded ball differs by some constant. Being unbounded, boundary dissipations do not matter in this scenario. In our present work, we analyze the transience class problem in the setting of bounded graphs. Indeed under the other assumptions we make about the graph families, it is the presence of sink node that makes the problem interesting and non-trivial. Our results basically show that the dissipation of sand particles through the boundary is limited (by a polynomial), consequently in polynomially many particle additions, the entire sandpile is flooded. \newline

\noindent\textbf{Diffusions and Potential Theory on Graphs:} Delmotte \cite{TD97} shows how Harnack's inequality leads to non-trivial results on heat diffusions. They talk about bounding the growth of harmonic functions using Gaussian estimates of the heat kernel, all of which follows from the assumption that the graph under consideration follows a parabolic Harnack's inequality. Chung and Yau \cite{CY94} derive the Harnack inequality for certain degree bound graphs. The bounds are then used to bound the Neumann eigenvalues of these graphs. In \cite{CY00}, they derive Harnack's inequality of Abelian homogenous graphs. Further more, Chung and Yau \cite{CY96} derive lower bounds on log-Sobolev constants by establishing log-Harnack inequalities on graphs. Bounding the log-Sobolev constants is important as it helps in establishing better convergence bounds on random walks on graphs, see Diaconis and Saloff-Coste \cite{DS96}. Chung and Yau \cite{CY95}, derive bounds on the eigenvalues of the Laplacians using the Sobolev inequalities and heat kernel estimates. \newline

\noindent\textbf{Electric Networks:} The classical theory of random walks (\cite{NW}, \cite{DS},\cite{Lov}) has some very  powerful and intuitive results which have recently found widespread application in theoretical computer science. Christiano, Kelner, M\k{a}dry and Spielman \cite{CKMS} have recently announced the fastest known algorithm for computing approximate maximum $s-t$ flows in capacitated undirected graphs. Their algorithm constructs the approximate flows by using the electric current flows on the same network with $s$ and $t$ as poles in an essential way. Earlier Kelner and M\k{a}dry \cite{KM} used arguments based on random walks to formulate the fastest known algorithm for generating spanning trees from uniform distribution. Spielman and Srivastava \cite{SS} construct good sparsifiers of weighted graphs via an efficient algorithm for computing approximate effective resistance between any two vertices, a result which is quite insightful on its own. Indeed a deeper understanding of harmonic functions is as much of interest to a computer scientist as to a potential theorist and as such the benefit of this confluence of different types of researchers has been mutual. A significant example is the, now seminal, work by Arora, Rao and Vazirani \cite{ARV} on embeddings of negative type which give an $O(\sqrt{\log n})$ approximation algorithm for computing graph conductance. \newline


\noindent\textbf{Other Results in Sandpiles:} As already mentioned, research problems in abelian sandpile model span across numerous areas. We make a passing mention to some of this work. Recent advances in complexity theoretic flavor include proof of the one-dimensional sandpiles prediction problem in \textbf{LOGDCFL} by Peter Bro Milterson \cite{Pet}. Schulz \cite{MS} mentions a related NP-complete problem. The group structure of the space of recurrent configurations, first introduced by Dhar, Ruelle, Sen and Verma in \cite{DD95} is also considered a fertile area of analysis. Cori and Rossin \cite{CR00} show that sandpile groups of dual planar graphs are isomorphic. Toumpakari \cite{Tou} discusses some interesting properties of sandpile groups of regular trees. Questions related to group rank are studied in particular, the paper is concluded with an interesting conjecture on the rank of all Sylow subgroups of the sandpile group. Specific families of graphs like square cycles $C^{2}_{n}$, $K_{3} \times C_{n}$, $3 \times n$ twisted bracelets, etc have been analyzed. We refer the reader to \cite{CHW}, \cite{SH}, \cite{HL}.

\end{subsection}

\section{Basic Properties of The Abelian Sandpile Model}

Our notation and terminology follows Babai and Gorodezky \cite{LB07}.

\begin{defin}A {\em{graph}} $G$ is an ordered pair $(V(G),E(G))$ where $V(G)$ is called the set of vertices and $E(G)$ is a set of $2-$subsets of $V$, possibly with repeated elements, the set of edges.\end{defin}

For brevity, even in the event of $G$ being a multi-graph, we will use graph. The number of edges in $E$ which contain $v$  is defined as the \textit{degree} of a vertex $v\in V$.  Two vertices $v$ and $u$ are called \textit{adjacent} (or neighboring) if $(u,v) \in E$. A path between two vertices $u$ and $v$ is an ordered sequence of edges $e_{1}, e_{2}, \ldots, e_{k}$ such that $u \in e_{1}$, $v \in e_{k}$ and for all values of $i$, $e_{i} \cap e_{i+1} \neq \phi$. The graph $G$ is \textit{connected} if there exists a path between any pair of vertices.

For an Abelian Sandpile Model, we take a connected graph $G$ with a special vertex called the \textit{sink}, denoted $s \in V$. We will be referring to the non-sink vertices in $G$ as ordinary vertices and denote this set by $V_{o} = V - \{s\}$.

\begin{defin} The {\em{configuration}} of a sandpile $G$ is a map, $c:V_{o} \rightarrow \mathbb{Z}^{+}$, which records the number of sand particles that each of the ordinary sites contains. It will be represented as a vector. The weight of $c$ is $|c| = \sum_{v \in V_{o}}c(v)$. \end{defin}

The \textit{empty} configuration is the zero vector. The \textit{capacity} of a site is the maximum number of particles that it can stably hold and is one less then the degree of the node.

\begin{defin} An ordinary node $v$ is said to be {\em{unstable}} in a configuration $c$ if $c(v) \geq \text{degree}(v)$. If all the sites in a configuration stable, the configuration is stable, else it is referred to as unstable. \end{defin}

When a site is unstable it is said to \textit{topple}, i.e. pass on some of its particles to its neighbors. When a site $v$ topples once, it loses $\text{degree}(v)$ particles and each neighbor of $v$ acquires a particle for every edge common with $v$. The sink node, by definition, never topples. We start with the empty configuration and keep adding particles one by one on sites of our choice and topple when necessary. The ASM evolves in time through these two modes, particle addition at sites and relaxation of unstable sites via toppling. A toppling sequence is an ordered set of configurations where every configuration can be obtained from the previous one by toppling some site unstable in it. The case of many sites becoming unstable simultaneously also poses no complication as the order in which they are subsequently relaxed does not effect the final stable configuration that is obtained at the end of toppling sequence, hence the prefix abelian. Elementary proofs of such confluence properties can be found in the pioneering paper on ASMs by Dhar \cite{DD90}.


\noindent \textbf{Notation:} We write $c_{1} \geq c_{2}$ if $\forall v, c_{1}(v) \geq c_{2}$ and $c_{1} \vdash c_{2}$ if there is a toppling sequence which takes $c_{1}$ to $c_{2}$. Lastly we write, $c_{1} \rightarrow c_{2}$ if $\exists c_{3} \geq c_{1}$ such that $c_{3} \vdash c_{2}$. We say that a configuration $c_{2}$ is \textit{reachable} from $c_{1}$ if $c_{1} \rightarrow c_{2}$ and \textit{unreachable} otherwise. In words, one can add particles to certain sites in $c_{1}$ so that there exists a toppling sequence leading to $c_{2}$. Note that being reachable is a transitive relation, i.e. $c_{1} \rightarrow c_{2}, c_{2} \rightarrow c_{3} \Rightarrow c_{1} \rightarrow c_{3}$.

\begin{thm}(\cite{DD90},\cite{auto}) Given any configuration $c$, there exists a unique stable configuration $\sigma(c)$ such that $c \vdash \sigma(c)$, independent of the toppling sequence chosen. \end{thm}

This makes sure that once we add some particles at some sites, the final state in well defined. Consequently, in the sequel, we will not be worrying about the configurations under discussion being stable/unstable. As long as any member of a toppling sequence is given, the entire sequence including the final stable configuration, is identified.

\begin{prop}\label{prop:linSupConf}If $c \vdash \sigma(c)$, then $kc \vdash k\sigma(c)$ \end{prop}

Property \ref{prop:linSupConf} will used extensively, later, for demonstrating impulse propagation in sandpile. Associated with every toppling sequence is the count of the number of times each site has toppled, the vector of \textit{toppling potentials}, also referred to as the \textit{score vector} in \cite{LB07}. These toppling potentials are very closely related to the electric potentials that develop at various nodes when power source-sink are appropriately applied, a connection which we will discuss in detail in the coming sections.

\begin{defin} Assuming $c_{1} \vdash c_{2}$, the toppling potential function $z^{c_{1}, c_{2}} : V_{0} \rightarrow \mathbb{Z}^{+}$ is defined as $z^{c_{1}, c_{2}}(v) :$ the number of times $v$ toppled in a toppling sequence from $c_{1}$ to $c_{2}$. We denote $z^{c, \sigma(c)}$ by $z^{c}$.\end{defin}

This function is well defined as the number of times a particular site topples is independent of the toppling sequence chosen, already noted in \cite{LB07}. A simple proof follows from typical linear algebraic arguments and the fact that the principal minor of a connected graph's combinatorial Laplacian is of full rank.

A configuration is called {\em{recurrent}} if it is reachable from \textit{any} configuration. As already mentioned, we say that a configuration $c_{i}$ is reachable from a configuration $c_{j}$ if by adding some particles to $c_{j}$ and subsequently relaxing it, we can obtain $c_{i}$. A configuration is {\em{transient}} if it is not recurrent. The set of recurrent configurations is therefore, closed under being reachable.

\begin{defin}A configuration $c$ is {\em{recurrent}} iff $\forall c'$ we have $c' \rightarrow c$.\end{defin}

Since recurrence persists under particle addition, we have the following property.

\begin{prop} If configuration $c_{1} \leq c_{2}$, then recurrence of $c_{1}$ implies that of $c_{2}$.\end{prop}

Denote the configuration in which every node $v$ has $\text{degree}(v)$ particles by $c_{max}$. Clearly, given any stable configuration $c$, one can reach $c_{max}$ simply by adding the required number of particles at each site.

We analyze the process of adding one grain at a time to the sandpile and study its evolution. As in the standard theory of Markov chains, recurrence characterizes the long term (steady state) behavior of sandpile. Our investigation is concerned with the maximum number of particles that can be added while staying transient. Following Babai and Toumpakari \cite{BT05}, for a sandpile $S$ we define the notion of transience class as follows.

\begin{defin}The {\em{transience class}} of $S$ denoted by tcl($S$), is defined as the maximum number of particles that can be added to $S$ before reaching a recurrent configuration.\end{defin}

\noindent \textit{Remark}: In \cite{CS11}, the authors talk about an alternate equivalent characterization of transience class which defines the transience class alternatively as \textit{the maximum number of particles that can be added before all the nodes have toppled at least once}. Even though we will not be using this definition explicitly, the notion is inherent in the way we bound transience classes. We will be computing the maximum number of particles that can be added at any point before at least one particle reaches every node. For sandpile with bounded node degrees, using the Property (\ref{prop:linSupConf}), this translates to the alternate characterization of transience classes we just mentioned (albeit with a constant multiplicative loss factor, the degree).


\section{Basic Properties}


Consider any (possibly infinite) graph $G(V,E)$. Distances in this graph will henceforth correspond to the shortest path metric. For any vertex $v$, $B(v,r)$ denotes the ball of radius $r$ around $v$, i.e. the set of all vertices in $G$ which are at a (shortest path) distance of at most $r$ from $v$. For any set of vertices $U \subseteq V$, define two notions of boundary

\begin{itemize}
 \item[-] vertex-boundary, $\delta_{V} U$ is the set of vertices, in $U$, having neighbors in the set $V - U$.
 \item[-] edge-boundary, $\delta_{E} U$ is the set of edges connecting vertices in $U$ to those in $V - U$.
\end{itemize}

For example if the set $U$ is the ball $B(v,r)$, the set $\delta_{V} B(v,r)$ is the set of vertices which are exactly at distance $r$ from $v$ and the set $\delta_{E} B(v,r)$ is the set of edges between the vertices in  $\delta_{V} B(v,r)$ and those in $\delta_{V} B(v,r+1)$. Consider any connected set of vertices $V_{h}$ and the induced subgraph $G_{h} = (V_{h}, E_{h})$. The sandpile corresponding to this sub-graph is obtained by adding the edge set $\delta_{E} V_{h}$ to $E_{h}$ and identifying all the vertices in $V - V_{h}$ which have neighborhood in $V_{h}$ with the sink node $s$. We denote this sandpile by $S = (V_{h} \cup \{s\}, E_{h}\cup \delta_{E} V_{h})$.

\begin{figure}[ht]
\centering
\includegraphics[width=.8\textwidth]{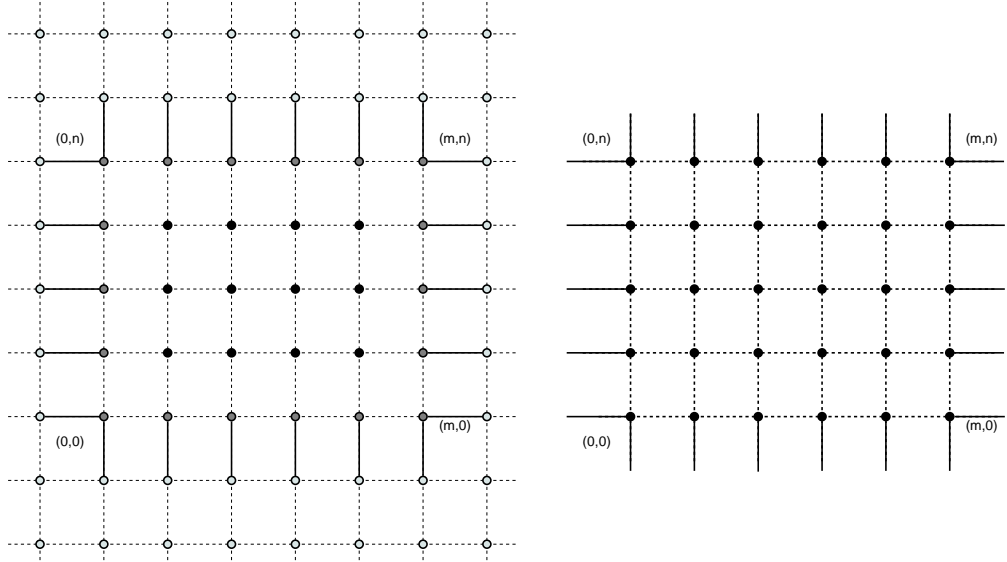}
\caption{A grid graph, the subset set $V_{m,n}$ and the corresponding sandpile}
\label{fig:ExGrid}
\end{figure}

\noindent \textbf{Example of grid sandpile:} To make the above definition more transparent, consider the example of grid sandpile. Consider the infinite grid graph with its canonical embedding in plane. For every pair of integers $(i,j)$ their exists a vertex (with this pair as its label). Each vertex $(i,j)$ is adjacent with $(i+1,j)$, $(i,j+1)$, $(i-1,j)$ and $(i,j-1)$, which are just the four lattice points flanking $(i,j)$. Let $V_{m,n}$ be the set of vertices with labels $\{ (i,j) : 0 \leq i \leq m, 0\leq j \leq n \}$. Evidently this set is connected. The boundary vertex set is the set of vertices lying on the horizontal line segments  $\{ (i,0) : 0 \leq i \leq m\}$ and $\{ (i,0) : 0 \leq i \leq m\}$ and on the vertical ones $\{ (0,j) : 0 \leq j \leq n \}$ and $\{ (m,j) : 0 \leq j \leq n \}$. The edge-boundary set is the set of edges  between vertices in the boundary set and the vertices not in the $(m+1)\times(n+1)$ block ($V_{h}$). Figure \ref{fig:ExGrid} exhibits the graph, the aforementioned boundary sets and the resulting sandpile. In the left figure, the white nodes belong to the set $V - V_{m,n}$. The grey nodes form the vertex boundary of $V_{m,n}$, and along with the black nodes constitute the set $V_{m,n}$. The heavy edges constitute the edge boundary. The figure on the right is the sandpile $S_{m,n}$. The heavy edges are connections to the special sink node.

We will denote both the sandpile and the underlying graph by the same symbol $S$. To keep notation and definitions clean, we fix the following convention. The ball $B(v,r)$ with respect to the shortest path metric is defined already. In the context of sandpile, however, when we say a ball $B(v,r)$ in some vertex subset $V_{h}$, or the induced sandpile $S$, we assume it is small enough to not contain the sink node. Also, whenever a sandpile is mentioned without explicit description of an ambient graph, we assume that a super-graph indeed exists and the sandpile has been obtained in the manner described above. Let $\mathcal{S} \equiv \{ S_{i}\}$ be an indexed family of sandpile graphs. We will now define some basic properties of these (possibly infinite) families which we will need later in discussion on bounding the transience class.

\begin{defin} Polynomial Volume Growth Property ($V_{\alpha}$) : A graph family $\mathcal{S}$ has polynomial volume growth property ($V_{\alpha}$) with growth exponent $\alpha$ if there exist constants $\delta$ and $\Delta$ such that for any graph $S_{i} \in \mathcal{S}$ and every node $v \in V(S_{i})$, if we consider the $B(v,r)$ around $v$, its volume satisfies the following bounds.

\begin{eqnarray}
 \delta.r^{\alpha} \leq \text{Vol}(B(v,r)) \leq \Delta.r^{\alpha}
\end{eqnarray}

Here, $\text{Vol}(B(v,r))$ is the number of edges inside the ball $B(v,r)$.
\end{defin}


This property ensures that there is a bound on the rate at which the measure grows. This becomes crucial if one wants to make any arguments along the lines of diffusions in Euclidean spaces. An easy consequence is the degree bound on the graph family $\mathcal{S}$. All vertices in every member graph have degrees bound between $\delta$ and $\Delta$.

\begin{defin} Degree boundedness ($\Delta$) : A graph family $\mathcal{S}$ satisfies ($\Delta$) if for every graph $S_{i} \in \mathcal{S}$, every node's degree is bounded from above by $\Delta$. Moreover, a sandpile satisfies ($\Delta$) if every normal (non-sink) node has degree bounded by $\Delta$.\end{defin}

If adding $x$ particles at some node $u$ causes some node $v$ to receive a particle at any point in time, we say that $v$ is \textit{flooded}. Similarly, we say that the set $V' \subset V$ got flooded if all the nodes, it consisted of, received at least one particle. There are two symmetry properties that will be important to us.

\begin{defin} High Local-Conductance Property (hLC($C_{\sigma}$)) : A sandpile family $\mathcal{S}$ satisfies hLC($C_{\sigma}$) if there exists a constant $C_{\sigma}$, independent of the index $i$ of the member sandpile $S_{i}\in \mathcal{S}$, such that for any site $v \in V(S_{i})$ and any ball $B(v,r)$ in $S_{i}$, placing at most $C_{\sigma}.\text{Vol}(B(v,r))$ particles at $v$ floods the ball $B(v,r)$. 
\end{defin}

This property implies that there is not too much dissipation of particles through the boundary of the sandpile, as long as we restrict ourselves to ensuring flooding balls that lie completely inside the set of ordinary vertices.


\begin{defin}\label{def:MV} Mean Value Property (MV($C_{h}$)) : A sandpile family $\mathcal{S}$ satisfies MV($C_{h}$) if there exists a constant $C_{h}$, independent of the index $i$ of the member sandpile $S_{i}\in \mathcal{S}$, such that for any member graph $S_{i} = (V_{h} \cup \{s\}, E_{h}\cup \delta_{E} V_{h})$ if we consider any function $\pi:V(S_{i}) \rightarrow [0,1]$ defined over $V(S_{i})$ which is harmonic over $V_{h}$, the following inequality holds for every ball $B(v,r)$ inside $V_{h}$,
\begin{eqnarray}
 \sum_{u \in B(v,r)}\pi(u) \geq C_{h}.\pi(v).\text{Vol}(B(v,r))
\end{eqnarray}
\end{defin}

The continuous analog of mean value properties occupy central position in classical analysis. See Section (\ref{sec:comCase}) for further discussion on the importance of this assumption. We now define an impulse superposition property.

\begin{defin} Local Superposition Property (LS($C_{l}$)) : A sandpile family $\mathcal{S}$ satisfies LS($C_{l}$) if there exists a constant $C_{l}$, independent of the index $i$ of the member sandpile $S_{i}\in \mathcal{S}$, such that given a ball $B(v,R) \subset S_{i}$ for any member graph $S_{i} \in \mathcal{S}$, if placing $H$ particles at $v$ topples a site $w$ in $B(v,R)$ and placing $h$ particles at every site in the smaller ball $B(v,r)$ also topples $w$, with $h$ minimum,
\begin{eqnarray}
 h \leq C_{l}.\frac{H}{\text{Vol}(B(v,r))}
\end{eqnarray}
\end{defin}

\noindent \textit{Remark:} This property essentially means that if placing $H$ particles at a site $v$ causes a toppling at site $w$, then the same effect can be produced at $w$ by adding a constant factor times $H$ particles distributed evenly among the sites in the ball $B(v,r)$. Such impulse distribution properties are easy to prove in the setting of classical complex analysis. The goal of proving a much more general superposition property, where the impulse is distributed not over a ball but a general set of nodes which are distance wise (in potential theoretic sense) well distributed with respect to the point of observation, seems unlikely to work out. There are simple counterexamples where such general distributivity does not work. However, it is also not known if the constraint of taking nodes inside a ball is really essential.


\begin{thm} \label{thm:polyVolSup} Given a sandpile family $\mathcal{S}$ which satisfies MV($C_{h}$) and ($\Delta$), it also satisfies LS($C_{l}$). Moreover, the constants $C_{l}, C_{h}$ and $\Delta$ satisfy the following relation.
\begin{eqnarray}
C_{l} = \frac{\Delta + 1}{C_{h}} \nonumber
\end{eqnarray}
\end{thm}


We defer the proof of this theorem to a later section.

\begin{defin} Overlapping Potentials Property (OP($f(.)$)) : Given a sandpile family $\mathcal{S}$. Consider a member $S_{i}$ and let $h$ be the smallest number of particles which when placed at every site in a ball $B(v,r) \subseteq S_{i}$, makes every site in the ball $B(v,R) \subseteq S_{i}$ topple at least once. If $h$ is bounded by a function $f(.)$ of $R/r$, independent of the sandpile index $i$, $\mathcal{S}$ satisfies the \textit{overlapping potentials property}, OP($f(.)$).
\end{defin}

\begin{lem}\label{lem:proofOP} Given a sandpile family $\mathcal{S}$ which satisfies hLC($C_{\sigma}$), MV($C_{h}$) and ($V_{\alpha}$), it also satisfies OP($f(.)$). Moreover, the function $f$ is of the following form.
\begin{eqnarray} \label{equ:OPpolynomial}
f(R/r) = \frac{C_{\sigma}}{C_{h}}.\frac{\Delta(\Delta + 1)}{\delta} \left( \frac{R}{r} \right)^{\alpha}
\end{eqnarray}
\end{lem}

\textbf{Proof:} Given a sandpile family $\mathcal{S}$ which satisfies hLC($C_{\sigma}$), MV($C_{h}$), and $V_{\alpha}$. Using theorem \ref{thm:polyVolSup}, it also satisfies LS($C_{l}$). Consider a member sandpile $S_{i} \in \mathcal{S}$. Consider any vertex $v \in V(S_{i})$ with the balls $B(v,r)$ and $B(v,R)$ around it with $R>r$. Let $x$ be the minimum number of particles one needs to place at $v$ to topple every site in $B(v,R)$ at least once. Since $S$ satisfies hLC($C_{\sigma}$), we have

\begin{eqnarray}
 x & \leq & C_{\sigma}.\Delta.R^{\alpha} \nonumber
\end{eqnarray}

The property ($V_{\alpha}$) implies the following lower bound on the volume of the ball $B(v,r)$.

\begin{eqnarray}
\text{Vol}(B(v,r)) & \geq & \delta.r^{\alpha} \nonumber
\end{eqnarray}

Now, consider any site $p$ in $\delta B(v,R)$. Using LS($C_{l}$), if $x$ particles placed at $v$ induce a toppling at $p$, then $C_{l}.x/\text{Vol}(B(v,r))$ particles at each of the sites in $B(v,r)$ necessarily induce a toppling at $p$. So, the minimum number of particles needed to be placed at each node of $B(v,r)$, to topple every site in $B(v,R)$ at least once, satisfy the following bound,

\begin{eqnarray}
h \leq C_{l}.C_{\sigma}.\frac{\Delta}{\delta} \left( \frac{R}{r} \right)^{\alpha} \nonumber
\end{eqnarray}

Again, using theorem \ref{thm:polyVolSup}, $C_{l} = (\Delta + 1)/C_{h}$. This gives us the final form of the polynomial.

\begin{eqnarray}
h \leq \frac{C_{\sigma}}{C_{h}}.\frac{\Delta(\Delta + 1)}{\delta} \left( \frac{R}{r} \right)^{\alpha} \nonumber
\end{eqnarray}

Therefore, $h$ is indeed bounded by a polynomial in the ratio $R/r$. $\square$

\noindent \textit{Notation:} With each site $v$ in a sandpile $S$, we associate the number $\eta(v) = \text{dist}(v, \delta_{V}(S))$ i.e. the distance between site $v$ and the vertex boundary of $S$. It is also the radius of the \textit{largest} ball centered at $v$ which is inside $S$.

\begin{figure}[ht]
\centering
\includegraphics[width=.8\textwidth]{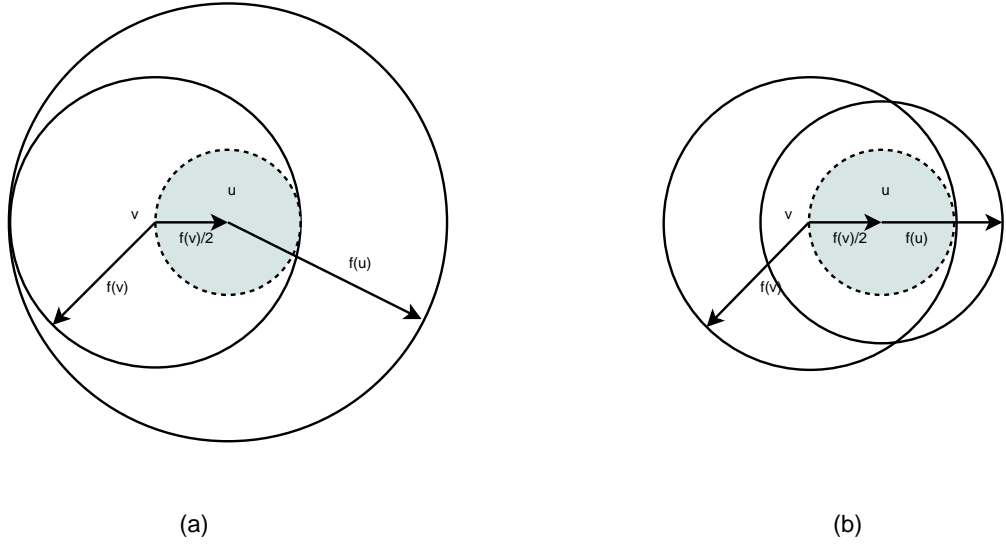}
\caption{Epicenter Propagation : Single step $(a)$ Expansion $(b)$ Contraction}
\label{fig:singStep}
\end{figure}

\begin{lem}(\textbf{Epicenter Propagation : Single Step})\label{lem:constFac} Given a sandpile family $\mathcal{S}$ which satisfies OP($f(.)$). Let $v$ be a node in $S_{i}\in\mathcal{S}$, with $B(v,\eta(v))$ as the largest ball around it. Let $u$ be any site in $\delta_{V} B(v,\lfloor \eta(v)/2) \rfloor$ (i.e. at distance $\lfloor \eta(v)/2 \rfloor$ from $v$). Then, there exists a constant $K$, independent of the sandpile index $i$, such that if a configuration $c$ floods the ball $B(v,\eta(v))$, then configuration $K.c$ floods $B(u,\eta(u))$.

Furthermore, if the sandpile family $\mathcal{S}$ satisfies hLC($C_{\sigma}$), MV($C_{h}$) and ($V_{\alpha}$), then the constant $K$ satisfies the following bound.
\begin{eqnarray}\label{equ:EpiPropCons}
K & \leq & \frac{C_{\sigma}}{C_{h}}.\frac{\Delta(\Delta + 1)}{\delta}.3^{\alpha}
\end{eqnarray}
\end{lem}

\textbf{Proof:} To keep the notation clean, we drop the floors. The argument without this simplification needs no new ideas and can be essentially reconstructed from the given proof.

In the sandpile $S_{i}\in\mathcal{S}$, consider any shortest path connecting $v$ to $u$, where $u \in \delta_{V} B(v, \eta(v)/2)$. The value of $\eta$ can vary by at most one with every step. Therefore, $\eta(v)/2 \leq \eta(u) \leq 3\eta(v)/2$. In particular, the ratio of radii is is bounded.

\begin{eqnarray}
 \frac{\eta(u)}{\eta(v)/2} & \leq & 3 \nonumber
\end{eqnarray}

Given the configuration $c$ which floods the ball $B(v, \eta(v))$. Since $B(u, \eta(v)/2) \subset B(v,\eta(v))$, $c$ also floods $B(u,\eta(v)/2)$. Using the overlapping potentials property (OP), the number of particles needed at every site in the ball $B(u, \eta(v)/2)$, to flood the normal ball $B(u,\eta(u))$, is bounded by a polynomial in $2\eta(u)/\eta(v)$. The equation (\ref{equ:OPpolynomial}) illustrates this polynomial. Putting in the appropriate value of radii, we get the following polynomial.

\begin{eqnarray}
\frac{C_{\sigma}}{C_{h}}.\frac{\Delta(\Delta + 1)}{\delta} \left( \frac{\eta(u)}{\eta(v)/2} \right)^{\alpha} \nonumber
\end{eqnarray}

We have already noted that the radius ratio in the above expression is bounded by $3$. Therefore, the constant $K$ is bounded by the following expression.

\begin{eqnarray}
K & \leq & \frac{C_{\sigma}}{C_{h}}.\frac{\Delta(\Delta + 1)}{\delta}.3^{\alpha} \nonumber
\end{eqnarray}

We call this constant $K$ and will be referring to this value in the coming sections. Using Property \ref{prop:linSupConf}, $K.c$ places $K$ particles on each site of $B(u, \eta(v)/2)$, which in turn necessarily floods $B(u, \eta(u))$. The expression in equation (\ref{equ:EpiPropCons}) is clearly independent of the sandpile index $i$. This completes the proof. $\square$

Note that there are two distinct case here, $\eta(u) \leq \eta(v)$ and $\eta(u)\geq \eta(v)$. Even though one argument suffices to obtain bounds for both the cases, they are qualitatively different. See figure (\ref{fig:singStep}). The case in which the distance from sink node is increasing along the path is termed the expansion step and the other case is called the contraction step. There is a third variant also, a drift step. The need for such distinction will become clear when we prove the general multi-step version of this lemma in the next section.

\noindent \textbf{Remark:} The above lemma allows one to shift the focal point of diffusion by some distance at the cost of an additional multiplicative constant. This generalizes the traveling diamond lemma of Babai and Gorodezky \cite{LB07}. This lemma forms the backbone of the main result. While proving the polynomial bounds on the transience class, we will be using this lemma to keep shifting the \textit{epicenter} from some starting node to a target node. We will bound the total number of applications of this lemma by $O(\log{n})$ thereby bounding the total particle requirement in $poly(n)$.

\section{Bounding the transience class}

In the proof of the Lemma (\ref{lem:constFac}) we use the fact that along any shortest path joining a vertex $v$ to a vertex in $B(v,r)$ (lying inside the sandpile), the distance from sink cannot increase or decrease too fast. For our purpose, this precondition of $u$ belonging to $B(v,r)$ is too restrictive. We will impose a more general constraint on the graph structure using the non-empty interior (NI) property for paths in graphs. Before that, we will need some basic definitions. If $P$ is path from $u$ to $v$, and $w \in P$ is some vertex on that path, then $\text{dist}_{P}(w, u)$ is the distance of $w$ from $u$ \textit{along} the path $P$. So, if $P = \{u_{0}. u_{1}, \ldots, u_{k} \}$ then $\text{dist}_{P}(u_{0}, u_{i})=i$.

\begin{defin} The function $\eta : V \rightarrow \mathbb{R}$ varies {\em{linearly}} over path $P$ (from $v_{1}$ to $v_{2}$) if for any site $v$ at a distance $\text{dist}_{P}(v_{1}, v)$ \textit{along} $P$ from $v_{1}$, we have the following two sided bounds on $\eta(v)$.
\begin{eqnarray}
 a_{l} + \eta(v_{1}) + b.\text{dist}_{P}(v_{1}, v) \leq \eta(v) \leq a_{u} + \eta(v_{1}) + b.\text{dist}_{P}(v_{1}, v) \nonumber
\end{eqnarray}

  Here, $a_{u}$, $a_{l}$ and $b$ are constants with absolute values bounded by $1$.
\end{defin}

\noindent \textit{Note}: The value of $b$ lies in the closed interval $[-1, 1]$ as the distance from sink cannot increase (or decrease) by more then one, when one step is taken along the path.

Consider a sandpile family $\mathcal{S}$ and a member graph $S_{i}$.  Call a path $P$ from $v_{1}$ to $v_{2}$ $(1,l)-$\textit{central} if $\eta(v_{i})$ values vary linearly over $P$, or if its length is bounded by $l\log(|S_{i}|)$. We call it $(k,l)-$\textit{central} if it is a juxtaposition of at most $k$ central paths, where $k$ and $l$ are constants independent of graph index (and size).

\begin{defin} Non-empty Interior Property (NI) : A sandpile family $\mathcal{S}$ has non-empty interior property, NI($k,l$), if for every member sandpile graph $S_{i} \in \mathcal{S}$ and if for every pair of vertices $v,w \in V(S_{i})$, there exists a path between them which is $(k,l)-$\textit{central}. Here, $k$ and $l$ are constants independent of the sandpile index $i$.
\end{defin}

\begin{defin}Polynomial Transience Class Property (pTcl) : We say that the transience class of a sandpile family $\mathcal{S}$ is polynomially bounded if, for any member sandpile $S_{i} \in \mathcal{S}$, adding at most polynomial (in sandpile volume $|S_{i}|$) particles at any site induces a toppling at every other site.
\end{defin}

The following lemma proves the polynomial bound on transience class of certain sandpile families. It is a generalization of the expansion and contraction phases while flooding a grid, as mentioned in Babai and Gorodezky \cite{LB07}.

\begin{thm}(\textbf{Epicenter Propagation Theorem} Given a sandpile family $\mathcal{S}$ which satisfies hLC($C_{\sigma}$), OP($f(.)$) and NI($(k,l)$). Then $\mathcal{S}$ satisfies pTcl.

Moreover, if $\mathcal{S}$ also satisfies MV($C_{h}$), and $V_{\alpha}$, then for any member sandpile $S_{i} \in \mathcal{S}$, the transience class satisfies the following bounds
\begin{eqnarray} \label{equ:tclBound}
\text{tcl}(S_{i}) & \leq & C_{\sigma}\Delta n^{k+ (2l+1)k.\log_{\hat{g}}(\frac{C_{\sigma}}{C_{h}}.\frac{\Delta(\Delta + 1)}{\delta}.3^{\alpha})}
\end{eqnarray}
Here, $n = |S_{i}|$
\end{thm}
\textbf{Proof :} First consider the case $k=1$. The general case will follow immediately by taking the $k^{th}$ power of particles required in the special case. Consider the sandpile $S_{i} \in \mathcal{S}$, and a pair of vertices $p$ and $q$ with a $(1,l)-$\textit{central} path $P$ from $p$ to $q$. Denote the vertices on the path $P$ as $v_{0} = p, v_{1}, \ldots , v_{n} = q$. The value of $f$ changes linearly from $\eta(p) = a$ to $\eta(q) \approx a + b.n$. There are three essentially different cases which can arise, $b>0$ and $b<0$, when $\eta$ varies linearly, and path length is logarithmic in graph size, when $\eta$ varies sub-linearly. We consider them individually.
\begin{enumerate}
\item Case ($b>0$) \textit{Expansion Phase} : Refer to figure (\ref{fig:expansionPhase}). Initialize by adding some constant number of particles so that the normal ball $B(v_{3},3)$ is flooded. Let this constant be $K_{0}$. We will now repeat the following process iteratively till $q$ receives a particle.

    Assume that we are currently at vertex $v_{i}$ and the normal ball $B(v_{i},\eta(v_{i}))$ is flooded. Consider the vertex $v_{j}$ with $j = i+ \eta(v_{i})/2$ and the normal balls  $B_{s} = B(v_{j}, \eta(v_{i})/2)$ and $B_{t} = B(v_{j}, \eta(v_{j}))$. $B_{s}$ was flooded in the last iteration. We use this to flood the concentric ball $B_{t}$. Using Lemma (\ref{lem:constFac}), if configuration $c$ flooded $B(v_{i}, \eta(v_{i}))$, then $K.c$ floods $B(v_{j}, \eta(v_{j}))$ for constant $K$. In every iteration, the radius of the target ball $B_{t}$ increases by at least a factor of $\eta(v_{j})/\eta(v_{i})$.

    \begin{eqnarray}
    \frac{\eta(v_{j})}{\eta(v_{i})} & = & \frac{a + b.j}{a + b.i} \nonumber
    \end{eqnarray}

    Since, $j = i+ \eta(v_{i})/2$ we get,

    \begin{eqnarray}
    \frac{\eta(v_{j})}{\eta(v_{i})} & = & \frac{a + b.i +  \eta(v_{i})/2}{a + b.i} \nonumber
    \end{eqnarray}

    Separating out into two summands, we get,

    \begin{eqnarray}
    \frac{\eta(v_{j})}{\eta(v_{i})} & = & \frac{a + b.i}{a + b.i} +\frac{\eta(v_{i})/2}{\eta(v_{i})} \nonumber\\
                                    & = & 1 + b/2 \nonumber
    \end{eqnarray}

    Let us define $g = 1 + b/2$. The path length is bounded from above by $n = |V(S_{i})|$. Therefore, the total number iterations is at most $\log_{g}(n)$. Each step contributes a bounded multiplicative factor to the total count of particles needed. This implies that the total number of particles needed, say $N$, is bounded by the following expression.

    \begin{eqnarray}\label{equ:tclBoundEx}
    N & \leq & K_{0}.K^{\log_{g}(n)} = K_{0}.n^{\log_{g}(K)}
    \end{eqnarray}

    \begin{figure}[ht]
    \centering
    \includegraphics[width=.5\textwidth]{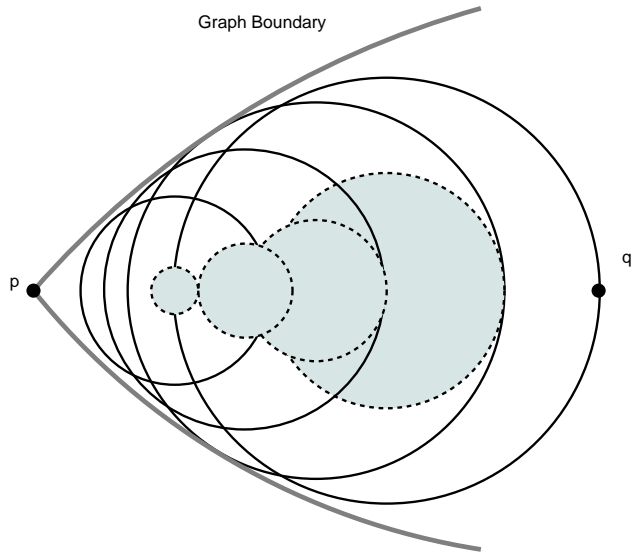}
    \caption{Epicenter Propagation : Expansion Phase - the radii of balls is expanding and the potential focus is moving towards the \textit{center} of graph interior}
    \label{fig:expansionPhase}
    \end{figure}

\item Case ($b<0$) \textit{Contraction Phase} : This case is analogous to the Expansion phase. See figure (\ref{fig:contractionPhase}). We initialize by flooding the ball $B(p, \eta(p))$. Using the property hLC($C_{\sigma}$), the number of particles required, say $K'_{0}$, is bounded by $C_{\sigma}.\text{Vol}(B(p, \eta(p)))$. But the volume of this ball is trivially bounded by the volume of the sandpile $S_{i}$, which in turn is trivially bounded by product of the maximum degree, $\Delta$, times the total number of vertices, $n$.

    \begin{eqnarray}
    K'_{0} & \leq & C_{\sigma}\Delta n \nonumber
    \end{eqnarray}

    We will now repeat the following process iteratively. Assume that we are currently at vertex $v_{i}$ and the normal ball $B(v_{i},\eta(v_{i}))$ is flooded. Consider the vertex $v_{j}$ with $j = i+ \eta(v_{i})/2$ and the normal balls  $B_{s} = B(v_{j}, \eta(v_{i})/2)$ and $B_{t} = B(v_{j}, \eta(v_{j}))$. In the last iteration, $B_{s}$ was flooded along with the ambient $B(v_{i},\eta(v_{i}))$. Using $B_{s}$, we will to flood the concentric ball $B_{t}$. Again, using Lemma (\ref{lem:constFac}), if configuration $c$ flooded $B(v_{i}, \eta(v_{i}))$, then $K.c$ floods $B(v_{j}, \eta(v_{j}))$ for some constant $K$. In every iteration, the radius of the target ball $B_{t}$ decreases by a factor of $\eta(v_{j})/\eta(v_{i})$.

    Following through the same computation as in the case of \textit{expansion phase}, we obtain that the contraction ratio is $1 + b/2$. Note that, since $b < 0$, this ratio is less than $1$. There is a constant factor \textit{contraction} in the radius of the balls flooded in each iteration. Let us denote the reciprocal of this ratio by $g' = (1 + b/2)^{-1}$. The implication of this remains the same though, that the total number of iterations are logarithmic in $n$. To be precise, the number of iterations is at most $\log_{g'}(n)$. Since each step contributes a multiplicative factor to the total particle requirement, the total number of particles needed, say $N$, is bounded by the following expression.

    \begin{eqnarray}\label{equ:tclBoundCon}
    N & \leq & K'_{0}.K^{\log_{g'}(n)} = K'_{0}.n^{\log_{g'}(K)}
    \end{eqnarray}

    \begin{figure}[ht]
    \centering
    \includegraphics[width=.5\textwidth]{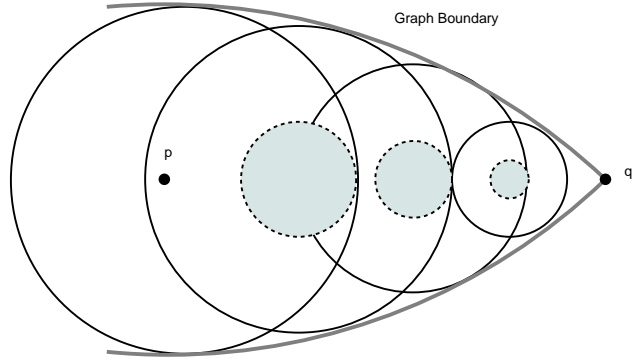}
    \caption{Epicenter Propagation : Contraction Phase - the radii of balls is going down and the potential focus is moving from the \textit{center} of graph interior towards the \textit{boundary} region}
    \label{fig:contractionPhase}
    \end{figure}

\item Case ($|P| = O(log(n)$)\textit{Lateral Drift Phase}: Let the path $P$ be such that $\eta$ varies sub-linearly over it. Also, let the minimum value of $\eta$ over $P$ be $\eta_{min}$. Consider the ball $B(p,\eta_{min})$. We flood this using at most $K'_{0} \leq C_{\sigma}\Delta n$ particles (follows from the (hLC($C_{\sigma}$)) property, just like the previous case).

    If configuration $c$ floods $B(p,\eta_{min})$, then $K.c$ floods $B(v_{\eta_{min}/2},\eta_{min})$, using Lemma (\ref{lem:constFac}). At a multiplicative cost of $K$, we cover a distance of $\eta_{min}/2$ on the path $P$. Let $l$ be a constant, independent of the sandpile index, such that $|P| \leq l.\log{n}$. We take at most $2l\log{n} / \eta_{min}$ steps in this fashion. See figure (\ref{fig:driftPhase}). The total number of particle required to finally flood the site $q$, say $N$, is bounded by the following expression.

    \begin{eqnarray}\label{equ:tclBounddrift}
    N & \leq & K'_{0}.K^{\frac{2l}{\eta_{min}}\log(n)} = K'_{0}.n^{\frac{2l}{\eta_{min}}\log(K)}
    \end{eqnarray}

    \begin{figure}[ht]
    \centering
    \includegraphics[width=.5\textwidth]{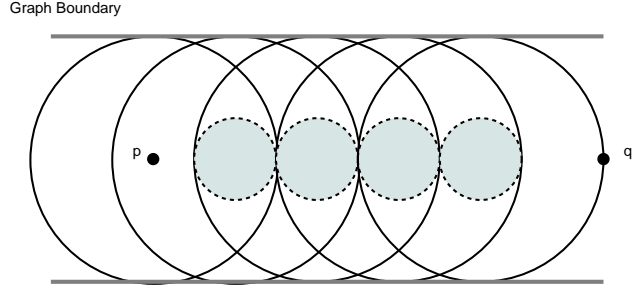}
    \caption{Epicenter Propagation : Lateral Drift Phase - the radii of balls under consideration stays same but the potential focus drifts inside the \textit{central} region of the graph from one focal point to another}
    \label{fig:driftPhase}
    \end{figure}
\end{enumerate}

Noting that $\eta_{min} \geq 1$, $K_{0} \leq K'_{0} = C_{\sigma}\Delta n$, and denoting $ \hat{g} = \min\{g,g'\}$, we have the following common bound on $N$ which respects the bounds in equations, \ref{equ:tclBoundCon}, \ref{equ:tclBounddrift}, and \ref{equ:tclBoundEx}.

\begin{eqnarray}
N \leq C_{\sigma}\Delta n^{1+ (2l+1).\log_{\hat{g}}(K)} \nonumber
\end{eqnarray}

After considering these cases of central paths, the argument of $k$-central paths follows directly. We split the path in the central components. Each of these components contributes a polynomial factor, mentioned above, in the total particles count. Since $k$ is a constant for the family of sandpile under consideration, we obtain that the total particle count is the following polynomial.

\begin{eqnarray}
N \leq C_{\sigma}\Delta n^{k+ (2l+1)k\log_{\hat{g}}(K)} \nonumber
\end{eqnarray}

Now assuming that the conditions $MV(C_{h})$, and $V_{\alpha}$ are also satisfied, and using bounds on the value of $K$ as mentioned in the Lemma (\ref{lem:constFac}), the bounds on $N$ can be written as,

\begin{eqnarray}
N \leq C_{\sigma}\Delta n^{k+ (2l+1)k.\log_{\hat{g}}(\frac{C_{\sigma}}{C_{h}}.\frac{\Delta(\Delta + 1)}{\delta}.3^{\alpha})}\nonumber
\end{eqnarray}


This completes the derivation of the theorem. \begin{flushright} $\square$\end{flushright}

The theorem demonstrates polynomial bounds on the transience class of certain sandpile families. Furthermore, it shows that the bounds are of the form,

\begin{eqnarray}
N \leq  k n^{p(\alpha)} \nonumber
\end{eqnarray}

where $k$ is a constant and $p(\alpha)$ is a linear function of $\alpha$, for the particular family. This form of bound can be contrasted with the Harnack's inequality, where the bounds on growth rates of harmonic functions look similar, with the dimension appearing in the exponent. Note that both the volume growth parameter $\alpha$ and the dimension of a Euclidean space, determine the measure contained in any region. Hence, the similarity confirms the intuition behind thinking of $\alpha$ as the dimension for certain kinds of graphs. 


\section{Impulse Superposition in sandpile : Proof of theorem \ref{thm:polyVolSup}}

We will now show that the local superposition property LS($C_{l}$) follows from mean value property MV($C_{h}$) and degree boundedness ($\Delta$) of graph. Let $H$ be the maximum number of particles which can be added at site $v$ without causing a toppling at site $w$. The following theorem from \cite{CS11} bounds the value of $H$ for degree bounded sandpile.

\begin{thm}(\cite{CS11}) For any sandpile with bounded vertex degrees, the minimum number of particles that need to be added at any vertex $v$ to observe a toppling at any vertex $w$ is equal, up to constant factors, to the following expression,
\begin{eqnarray}
\frac{1}{\pi_{w}(v)} \sum_{u}\pi_{w}(u)\nonumber
\end{eqnarray}
In particular, the bounds are,
\begin{eqnarray} \label{equ:boundH}
\frac{1}{(\Delta + 1)\pi_{w}(v)} \sum_{u}\pi_{w}(u) \leq H \leq \frac{(\Delta - 1)}{\pi_{w}(v)} \sum_{u}\pi_{w}(u)
\end{eqnarray}
\end{thm}



Similarly, let $h$ be maximum number of particles which when placed at each site of the ball $B(v,r)$ do not topple $w$. Let $z(u)$ denote the number of times the node $u$ topples in the process. Once the sandpile is stable, we observe that the number of particles on any particular node $u$ is between $0$ and $\text{degree}(u)-1$. This number is equal to the difference of total influx and outflow. The influx is due to neighbors' toppling and direct particle additions at $u$, and the outflow is due the toppling of node $u$ itself. Writing out the equations for each of these nodes and taking the rational relaxation gives us the following LP, which bounds the value of $h$ from above.


\begin{eqnarray}
\text{max } h \nonumber \\
  \forall u \in B(v,r) : 0 \leq \sum_{u' \sim u} z(u') - \text{degree}(u).z(u)  + h & \leq & \text{degree}(u) - 1\nonumber \\
  \forall u \notin B(v,r) : 0 \leq \sum_{u' \sim u} z(u') - \text{degree}(u).z(u)  & \leq & \text{degree}(u) - 1\nonumber \\
  z(w) \leq 0, z \geq 0, h \geq 0 \nonumber
\end{eqnarray}

There are two sets of flow conservation equations here, one for the nodes inside $B(v,r)$ where particles have been added directly, and the second for the remaining nodes in $B(v,R)-B(v,r)$. From the weak duality for LPs, it follows that to obtain an upper bound of $\alpha$ on the optimum value of the above system, it suffices to find a feasible solution of the dual LP of value $\alpha$. The following minimization program is the dual of the above.

\begin{eqnarray}
\text{min } \sum_{u}(\text{degree}(u)-1).Y(u)  \nonumber \\
\sum_{u' \sim w} Y(u') + Y' - \text{degree}(w).Y(w) & \geq & 0 \nonumber \\
\forall u \neq w : \sum_{u' \sim u} Y(u') - \text{degree}(u).Y(u) & \geq & 0 \nonumber \\
 \sum_{u \in B(v,r)} Y(u) & \geq & 1 \nonumber \\
  Y \geq 0, Y' \geq 0  \nonumber
\end{eqnarray}

Now, consider the following set of equations:

\begin{eqnarray}
\sum_{u' \sim w} Y(u') + Y' - \text{degree}(w).Y(w) & = & 0 \nonumber \\
\forall u \neq w : \sum_{u' \sim u} Y(u') - \text{degree}(u).Y(u) & = & 0 \nonumber \\
 \sum_{u \in B(v,r)} Y(u) & = & 1 \nonumber
\end{eqnarray}

A non-negative set of values satisfying the above set is feasible for the dual LP. We find these by considering the resistive circuit $\widehat{S}$, obtained by replacing each edge in $S$ by a unit resistance. We assign ground potential to the sink, and inject current at node $w$ such that it gets unit potential. The potential that develops on any node $u$ is $_{s}\pi_{w}(u)$. Evidently, all these potential values belong to the unit interval $[0,1]$. The sum of potential values at nodes in $B(v,r)$, $\sum_{u\in B(v,r)}\pi_{w}(u)$, can be used to scale the input current at $w$ thereby scaling all the potentials as well, such that the sum of potentials over the set of nodes $u \in B(v,r)$ becomes unit. It follows that the values $Y(u) = \pi_{w}(u)/\sum_{u\in B(v,r)}\pi_{w}(u)$ and $Y'$ equaling the value of the current injected form a feasible solution of the dual LP. This gives the following bound on $h$ in terms of the objective value at this point.

\begin{eqnarray}\label{equ:boundh-1}
h & \leq & \frac{1}{\sum_{u\in B(v,r)}\pi_{w}(u)} \sum_{u}(\text{degree}(u)-1).\pi_{w}(u) 
\end{eqnarray}

We are given that the sandpile graph satisfies the mean value MV($C_{h}$) property (definition (\ref{def:MV})). We restate the mean value inequality below.

\begin{eqnarray}\label{equ:MVS}
 \sum_{u \in B(v,r)}\pi_{w}(u) & \geq & C_{h}.\pi_{w}(v).\text{Vol}(B(v,r))
\end{eqnarray}

for some constant $C_{h}$ independent of the sandpile index. Using equations \ref{equ:boundh-1} and \ref{equ:MVS}, we get the following bounds on $h$ (up to constant factors).

\begin{eqnarray}\label{equ:boundh}
h & \leq & \frac{1}{C_{h}.\text{Vol}(B(v,r))}\frac{1}{\pi_{w}(v)}\sum_{u}(d(u)-1).\pi_{w}(u)
\end{eqnarray}

Now, using the lower bounds in equation \ref{equ:boundH}, which tightly approximates the value of $H$, and equation \ref{equ:boundh}, which bounds the value of $h$, we can bound $h$ in terms of $H$. The following relation is obtained.

\begin{eqnarray}\label{equ:boundhH}
h & \leq & \frac{\Delta + 1}{C_{h}}\frac{H}{\text{Vol}(B(v,r))}
\end{eqnarray}

The equation demonstrates the local superposition property LS($C_{l}$). In particular, it shows that the constant $C_{l}$ of LS($C_{l}$) is related to the constants $C_{h}$, of MV($C_{h}$), and $\Delta$, of ($\Delta$), in the following form.

\begin{eqnarray}
C_{l} & = & \frac{\Delta + 1}{C_{h}} \nonumber
\end{eqnarray}

This completes the proof of theorem \ref{thm:polyVolSup}.

\section{Special case of grid sandpile}

Figure (\ref{fig:ExGrid}) shows an example of an $m \times n$ grid based sandpile. We consider the family of sandpiles consisting of the symmetric case ($m=n$).

\begin{figure}[ht]
\centering
\includegraphics[width=.4\textwidth]{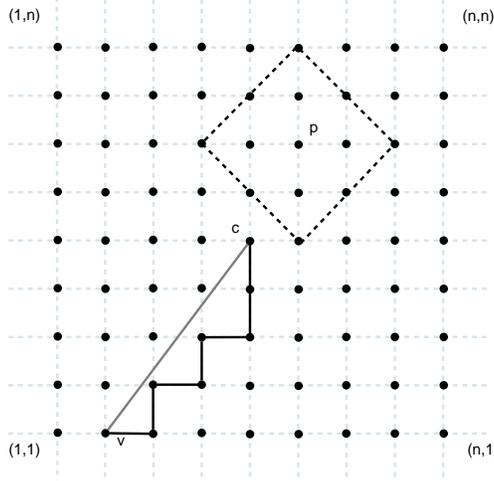}
\caption{An $n \times n$ grid sandpile : $\text{GRID}_{n}$}
\label{fig:grid}
\end{figure}

Consider an $n \times n$ grid graph. Attach an extra sink node $s$ to all the nodes lying on the boundary of this grid such that there is a double edge with every corner node and single edge for others. We denote the sandpile and graph by $\text{GRID}_{n}$. Figure (\ref{fig:grid}) shows an example.


As in the example of grid sandpile, we assume that our sandpile is coming from the induced subgraph over a suitable set of vertices in the infinite lattice. The reason behind this choice is that we will be talking about the \textit{family} of all grid sandpiles. While a particular finite grid can be a subgraph of many infinite graphs, the infinite grid is the canonical choice of a common ancestor of all finite grid graphs. We will show in this section that the infinite grid, $\mathbb{Z}^{2}$, along with the finite subgraph $\text{GRID}_{n}$ satisfy all the properties required to imply polynomial bounds on the transience class of grids.

\noindent \textbf{Degree bound ($\Delta$)}: The infinite grid  $\mathbb{Z}^{2}$ is regular with degree $4$. Every induced subgraph obeys this degree bound.

\noindent \textbf{Polynomial volume growth property($V_{\alpha}$)}: For any vertex $p$ and radius $r$, volume of the ball $B(p,r)$ grows as a quadratic in $r$. That is to say $\mathbb{Z}^{2}$ satisfies $V_{\alpha}$ with $\alpha = 2$. This simple fact can be proved using elementary counting arguments. Figure (\ref{fig:grid}) shows a typical ball around vertex $p$.

\noindent \textbf{Non-empty interior property(NI$(2,0)$)}: Given any finite grid, $\text{GRID}_{n}$, it suffices to show that there exists a path from center to every other node on which the distance to sink varies linearly. For simplicity, assume that $n$ is odd and take any point $v$ on the boundary. Denote the center node by $c$. Consider the canonical embedding of this grid in the plane. Join $c$ to $v$  by a line segment. Take the projection of this line segment towards $x-$axis and construct the path using the set of highest lattice points lying below the segment. This path has the requisite property and the growth parameter is just the slope of the line joining $c$ and $v$. Figure (\ref{fig:grid}) shows one such path. For an arbitrary pair of points, $u$, and $v$, the path is a juxtaposition of the paths from $u$ to $c$, and from $c$ to $v$. So at most two $(1,0)$-central components suffice. The argument for the more general case of $v$ lying inside the grid is identical. The case of even $n$ can be handled using essentially the same ideas.

\noindent \textbf{Mean Value property (MV)}: For a proof of the mean value property (MV), the reader is referred to Lemma 6 in \cite{JLS}. We reproduce it here for reference.

\begin{lem}(Exact Mean Value property on an approximate ball, \cite{JLS}) For each real number $r>0$, there is a function $w_{r}:\mathbb{Z}^{d}\rightarrow \left[ 0,1 \right]$ such that,
\begin{itemize}
\item $w_{r}(x) = 1$ for all $x \in \textbf{B}_{r-c}$, for a constant $c$ depending only on $d$.
\item $w_{r}(x) = 0$ for all $x \not\in \textbf{B}_{r}$.
\item For any function $u$ that is discrete harmonic on $\textbf{B}_{r}$,
    \begin{eqnarray}
    \sum_{x\in \mathbb{Z}^{d}}w_{r}(x)(u(x) - u(0)) = 0. \nonumber
    \end{eqnarray}
\end{itemize}
\end{lem}

This lemma allows us to get up to a constant distance of the boundary of the ball under consideration. The additional cost of sites in the annulus $\textbf{B}_{r-c,r}$ can be accounted for by loosing at most a constant factor, there by giving us the weaker inequality mentioned in the definition of property MV.

\noindent \textbf{high local-conductance (hLC)}: Babai and Gorodezky \cite{LB07} prove both (hLC) and (LS) property on grids, using elementary combinatorial arguments. We reproduce the discussion below without any essential changes. Note that the mean-value property is used in the Theorem \ref{thm:polyVolSup}, where it along with the property $\Delta$, implies the Local Superposition (LS) property. This is used in the proof of the Lemma \ref{lem:proofOP} to prove the Overlapping Potentials (OP) property. So, showing (LS) directly obviates the need to prove (MV) altogether.

\begin{defin}\textit{$D_{4}$ symmetry}: A function $f$ defined on an infinite two dimensional lattice, $\mathbb{Z}^{2}$, $f : \mathbb{Z}^{2} \rightarrow \mathbb{N}$ is said to have $D_{4}$ symmetry with respect to vertex $v$ if is symmetric with respect to all the axis of symmetry passing through $v$.
\end{defin}

In case of functions which are defined on a finite $n \times n$ grid, the above definition is applied by assigning zero to all the points on which the function is not defined. This creates a function with a finite support. The point $v$, in these cases, is bound to be the geometric center of this finite grid.

\begin{defin}\textit{Axis Monotonicity}: Let $f$ be a function defined on an infinite two dimensional lattice, $\mathbb{Z}^{2}$, $f : \mathbb{Z}^{2} \rightarrow \mathbb{N}$. We say $f$ is axis monotone about some vertex $v$ if $f(p) \leq f(q)$ for any pair of lattice points $p$ and $q$ such that the segment $q-p$ is aligned perpendicular to some axis of symmetry passing through $v$ and $q$ is closer to this axis then $p$.
\end{defin}

\begin{figure}[ht]
\centering
\includegraphics[width=.3\textwidth]{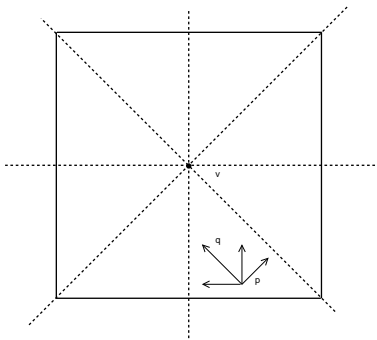}
\caption{Axis Monotonicity of a function}
\label{fig:axisMon}
\end{figure}

See figure (\ref{fig:axisMon}) for a illustration of directions of the axis-monotonicity and $D_{4}$ symmetry.

\begin{lem}\cite{LB07} If the starting (possibly unstable) configuration has $D_{4}$ symmetry and is axis-monotone, then so is the function $f : \mathbb{Z}^{2} \rightarrow \mathbb{N}$, where $f(v)$ is the number of times the site $v$ topples.\end{lem}

In the present context, the function of interest will be the number of particles present at a particular node. We observe that any function which is $D_{4}$ symmetric, axis-monotone, and has finite support, is sandwiched inside an $l_{1}$ ball (diamond) and $l_{\infty}$ ball (square) of same radius. The number of sites contained in both these balls is quadratic in the radius. To be precise, denoting an $l_{\infty}$ ball of radius $n$ around $v$ by $B_{\infty}(v,n)$, the volume of this ball is the maximum of particles it can hold in any stable configuration. This is given by,

\begin{eqnarray}
\mid B_{\infty}(v,n) \mid = 3(2n+1)^{2} \nonumber
\end{eqnarray}

Similarly, the maximum weight of any stable configuration on $B_{\infty}(v,n)$ is bounded by the following.

\begin{eqnarray}
\mid B_{1}(v,n) \mid = 6n^{2} + 6n + 3 \nonumber
\end{eqnarray}

The balls, therefore can hold at most $\Theta(n^{2})$ particles. When one starts adding particles at node $v$, clearly adding $\mid B_{1}(v,n) \mid = \Theta(n^{2})$ particles is enough to flood an  $l_{1}$ ball of radius $n$ around $v$. The property (hLC) follows immediately.

\begin{figure}[ht]
\centering
\includegraphics[width=.3\textwidth]{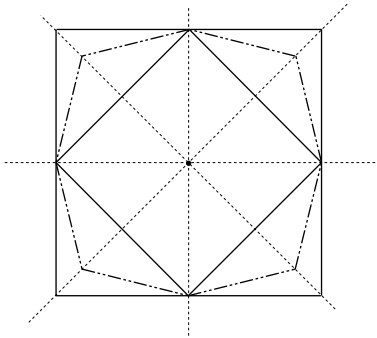}
\caption{A example of a function support which is $D_{4}$-symmetric and axis monotone (about the center of the grid). The function support is shown in broken lines and bold lines depict the $l_{\infty}$ and $l_{1}$ balls}
\label{fig:confSand}
\end{figure}


The property (LS) requires more work. Here we start with placing some $h$ particles at each node in $B_{1}(v,r)$. This configuration is axis-monotone and $D_{4}$-symmetric. Consequently, the stable configuration resulting from this is also axis-monotone and $D_{4}$-symmetric. Let $h$ be smallest such number such that the resulting configuration contains the ball $B_{1}(v,R)$. The total number of particles this configuration can have is $\Theta(R^{2})$. The particles we started of with are $h.\Theta(r^{2})$. Hence, $h$ is $\Theta((R/r)^{2})$. The property (LS) follows.

\section{Parallels with potential theory in continuous spaces}\label{sec:comCase}

We refer the reader to the appendices \ref{app:harmDisc}, and \ref{app:harmCont} for an introductory discussion on the theory of harmonic functions both in the discrete setting, graphs, and the continuous setting of real and complex spaces. In this section we will assume the basics mentioned in the appendices.

In analogy with the graphs induced on the vertex set $V_{h}$ and the boundary sets connecting the poles to interior, we talk about open sets in $\mathbb{R}^{n}$ and the boundary of these sets. Consider an open set $\Omega \subset \mathbb{R}^{n}$ and its boundary set $\delta \Omega$. Consider a positive function $h : \Omega \rightarrow \mathbb{R}^{+}$, that is \textit{harmonic} over the region $\Omega$. The reader is referred to appendix \ref{app:harmCont} for a formal definition and discussion on harmonic functions. The property of interest to us is elucidated by the celebrated Harnack's inequality.

\textit{Harnack's inequality for balls:} If $h$ is positive harmonic function defined over $\Omega$ then the following inequality holds for any (open) ball $B(p,r) \subseteq \Omega$ and $q \in B(p,r)$.

\begin{eqnarray}\label{equ:harnBall-main}
 h(q) \leq \left( \frac{r}{r - |p - q|} \right)^{2}h(p)
\end{eqnarray}

\begin{figure}[ht]
\centering
\includegraphics[width=.5\textwidth]{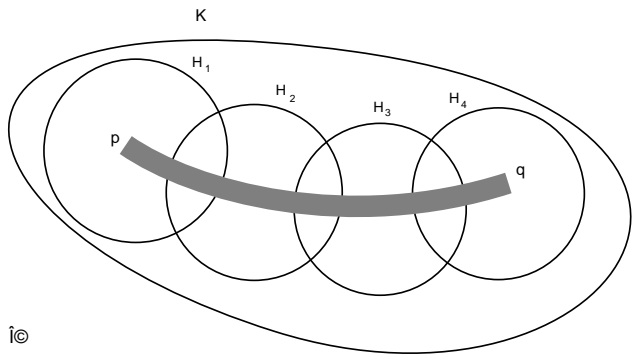}
\caption{A region $\Omega$, the path between $p$ and $q$ covered by balls}
\label{fig:harnCov-main}
\end{figure}

The generalization to higher dimensions is straightforward. From this, the inequality for all compact connected regions $K ( \subseteq \Omega)$ follows using a very simple chaining argument which involves considering a set of open balls in $K$ which cover the path from the reference vertex $p$ to the vertex $q$. Using the bounds from equation (\ref{equ:harnBall-main}), the growth rate of harmonic functions in each of these balls is bounded by the respective constants $H_{i}$. Multiplying these gives us an obvious bound over the considered path. By taking any finite covering of $K$ (compactness of $K$ guarantees the existence of such a covering), one can find an absolute bound on the value of the Harnack's constant by simply multiplying the constants corresponding to each of the balls (see figure (\ref{fig:harnCov-main})). This implies the Harnack's inequality over any compact region.

\textit{Harnack's inequality over general regions:} Suppose that $\Omega$ is connected and $K$ is a compact subset. Then for any harmonic function, $h$, defined over $\Omega$, the following inequality holds

\begin{eqnarray}\label{equ:harnComp-main}
 h(q) \leq H_{K} h(p)
\end{eqnarray}

Here $H_{K}$ is the Harnack's constant of the region $K$ and is independent of the function $h$. The Harnack's constant, in a sense, summarizes the connectivity of the corresponding region.

\begin{figure}[ht]
\centering
\includegraphics[width=.5\textwidth]{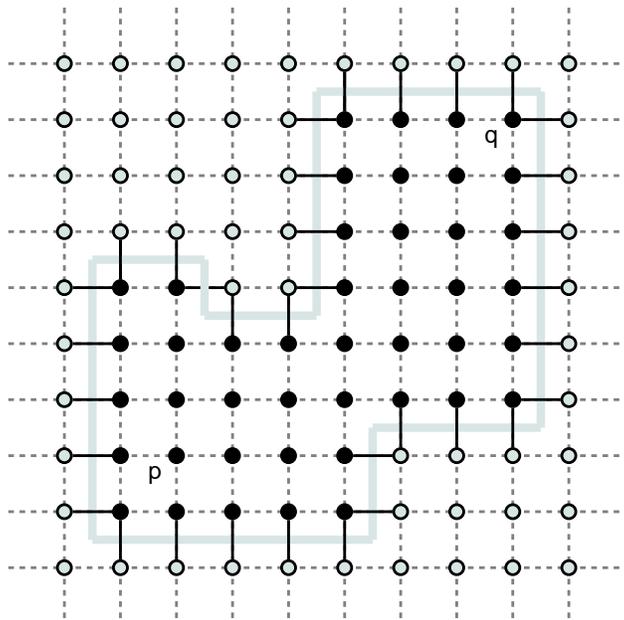}
\caption{A finite subset of the grid graph}
\label{fig:discHarn-main}
\end{figure}

Now consider the analogous situation in graphs. Consider the infinite grid graph, denoted by $\mathbb{Z}^{2}$. Consider any finite subset of vertices $V_{h}$ and the subgraph induced on them, in the sense discussed in previous sections. See figure (\ref{fig:discHarn-main}). The black nodes belong to $V_{h}$, heavy edges are incident on sink and for clarity the selected sub-grid is demarcated by the light thick line. The Harnack's constant for this region should measure the maximum possible growth in magnitude of a function which is harmonic over this domain. This is equivalent to finding the maximum value of $\pi_{p}(q)^{-1}$, with the sink node as the zero potential, and both $p$ and $q$ on the vertex boundary. Consider one such pair $p$ and $q$ (as shown in figure (\ref{fig:discHarn-main})). The value $\pi_{p}(q)$ measures the probability of a walk, starting at $p$, \textit{finding} $q$ before it dissipates out of the region. A high value (bounded from below by some constant, say) would imply that the nodes under consideration enjoy overall good connectivity and even a memory less strategy can find some path between them. A high value of $\pi_{p}(q)^{-1}$, on the the other hand, would imply that the pair is not so well connected in the potential theoretic sense and it is highly likely for a walk starting at $p$ to exit the system before hitting $q$ at all. Bounding the value of $\pi_{p}(q)^{-1}$ for all possibilities of $p$ and $q$ implies that even the pair with the worst case connectivity are not very bad. These properties are analogous to the mixing times of graphs without boundaries.

\noindent \textit{Remark}: The Figure (\ref{fig:harnCov-main}) shows the path from $p$ to $q$ covered by a set of compact balls. One can see that even if these balls have significant overlap, the discussion in preceding paragraphs is oblivious to the utility. One might as well have a set of balls, any pair of which intersects at most at one point. Evidently, the point to point connectivity, Brownian motion based diffusion type, is very different for the two cases and when the covering balls have significant overlapping, the Harnack's constant may be smaller. It therefore seems desirable to formulate properties which would allow better gluing of Harmonic functions leading to better bounds. \textit{For the continuous case, no such properties are known}. The triangle inequality for potentials (Lemma \ref{lem:potTri}) on graphs is the discrete version of the analogous property for continuous spaces is almost as old as the classical potential theory itself. This is enough to perform the stringing of Harnack's constants mentioned above. In the context of graphs, when we try to bound the transience classes, relying on this property alone clearly doesn't suffice. One can show that, in case of an $n \times n$ grid, about $\Theta (\log n)$ balls will be required to cover the path between opposite corners. Multiplying their transience classes yields super-polynomial bounds. The impulse superposition properties we derive, in the context of sandpile, are essential to show efficient bounds and form a completely new component in any effort towards demonstrating gluing properties of potential functions.


In the preceding sections, we demonstrate polynomial bounds on the transience class when the graph satisfies certain assumptions. We now discuss these assumptions, and try justifying these by showing analogous constraints in the continuous framework of harmonic functions.

\begin{itemize}
 \item[-] Volume growth parameter, $\alpha$, ensures uniformity in measure distribution throughout the space. Its role is analogous to the dimension of spaces in classical analysis. Whether we consider the counting measure associated with graphs and or the Lebasgue measure associated with continuous spaces, volume growths in both the cases are polynomials with exponents $\alpha$, and is uniform throughout the space under consideration. In the derivation of our bound on the transience class and consequently on the Harnack's constant of the graph, this analogy is evident.

 \item[-] Mean Value Property (MV) of graphs is similar to the mean value property of continuous spaces. This assumption underlines a certain homogeneity of space and uniformity in measure distribution. The main implication of such homogeneity is that a random walk which starts at center of a ball, induces a measure on the surface of this ball, which is uniform in the sense that the ratio of highest to lowest probabilities is bounded irrespective of the radius of the ball.

 \item[-] Non-Empty Interior Property (NI) is an analogue of the assumption of non-degeneracy of domains in analysis. This means that for every point inside the region, there is a full-dimensional ball containing it which lies completely inside the region. The concluding section also discusses degeneracy in graphs.

 \item[-] High local Conductance (hLC) is analogous to ensuring that the sandpile diffusion in space is almost isotropic. This means that when diffusion starts from any point, for any potential value, the distances of nearest and farthest equipotential points from the center are not much different. In continuous spaces, this is a simple consequence of symmetry. In fact, one can show much stronger property of conformal invariance, in the context of Brownian motions.

 \item[-] The Polynomial Transience Property (pT) is similar to ensuring that the Harnack's constant is not too large. Following from our previous discussion, this implies that the diffusion inside the region is reasonably good. To be precise, polynomial bounds on transience class mean that the corresponding sandpile achieves recurrence in reasonable time, and therefore is a good model of diffusion. Conversely, a sandpile with exponential transience class should correspond to a system which has not been modeled well.
\end{itemize}

%
%

\section{Conclusion}

In the present work, we generalize the results of Babai and Gorodezky \cite{LB07} beyond the family of grid graphs. The analysis raises lot of interesting questions, answers to which will aid our understanding of discrete diffusions.

The first question is, how far are we from a complete characterization? The present result shows that under some general and reasonable assumptions, the polynomial bounds can be established. There are some obvious extensions possible from our theorems. For example if one considers graphs made from attaching two graphs each individually satisfying the required pre-conditions but having different volume growth parameters. One can bound the Harnack's constant for both components separately and bound the one for the whole graph using the simple triangle inequality. But apart from these trivial modifications, we do not yet know of any class of sandpile which are not covered by our result. Supported by the analogy with the continuous case, we believe the assumptions we make are essential and can not be substantially relaxed.

The second question supplements the first one. Consider any Euclidean space and triangulate it using simplices or cubic complexes, basically any regular tessellation will do. We are interested in the graph of this triangulation complex. Indeed, the grid sandpile, which started of the study of transience class bounds, are obtained in exactly this manner from the two dimensional Euclidean space. In this graph, consider a subgraph corresponding to any open set. To be precise, the mesh size is also a parameter. So for every set in space, there exists a family of subgraphs parameterized by the fine-ness of triangulation. For example, in the case of grid sandpile, the underlying set is a square and the parameter $n$ in the $n \times n$ grid sandpile encapsulates the precision with which the discrete version can model the continuous version.

In this setting, the polynomial bounds in sandpile settings correspond to non-trivial bounds on the Harnack's constant in the continuous space. Any degenerate set in $\mathbb{R}^{n}$ has unbounded Harnack's constant. It can be shown that the sandpile corresponding to these sets also have transience classes which cannot be polynomially bounded. Babai and Gorodezky \cite{LB07} first mention this for line sandpile. The case of general degenerate sets is similar. The only crucial component is the fact that from every node, there is an edge to sink node. One can see an interesting confluence of the notion of bad models here. In the setting of topological spaces, full dimensional sets are considered to be \textit{available} or good models (in context of physics) while the degenerate ones are considered bad models. On similar lines, Babai and Toumpakari \cite{BT05}, and Babai and Gorodezky \cite{LB07} make case for the notion of polynomially bound transience classes as the essential qualification for a sandpile to be a good model. Dhar \cite{DD06} mentions several physical phenomena which sandpiles are intended to model. In all these settings, the sandpile graphs arise as tessellation graphs of the underlying continuous space. It is only natural to expect that if the underlying set itself is a bad model, the sandpile obtained on top of it is going to a bad model.

The line sandpile is a simplest case of this observation. This sandpile is based on a line graph which has two connections to sink from every node, except the corner ones which have three connections each. One can think of these sandpile as $n \times 1$ grids based sandpile (or degenerate grids, to make the connection sound clearer). Babai, Gorodezky, and Shapiro \cite{BGS} show exponential lower bounds on the transience classes of these sandpile using very simple combinatorial arguments. Independently, Choure and Vishwanathan \cite{CS11} also demonstrate these same bounds, using potential theoretic arguments. With some work, these can be extended to other degenerate cases. This categorically classifies the line sandpile as an infeasible model for diffusion. Now consider the set in $\mathbb{R}^{2}$ whose tessellation gives rise to such graph. The underlying set is a line segment in a plane, clearly a degenerate set. In this sense, \textbf{line sandpile} is a misnomer. The correct sandpile to associate with this name should be the one corresponding to a line segment in $\mathbb{R}$. This one has just two single connections to sink node, one at each end, and not surprisingly, has polynomially bound transience class. In fact any sandpile family of $k \times n$ grids, where $k$ is a constant, cannot correspond to an open set and should have exponential transience classes. Empirical results support this view. This leads us to a meta-conjecture of sorts.

\noindent \textbf{Question:} \textit{Does every bad sandpile correspond to a degenerate set?}

The constant factor approximation which we discuss in this paper, demonstrates that studying the transience class is similar to bounding the harmonic functions over the underlying graphs. The growth rates of harmonic functions in continuous spaces is a function of the dimension of the space. It is desirable to generalize this notion to graphs. In our present work, we are able to demonstrate bound on the Harnack's constant which is a function of the volume growth parameter, $\alpha$. However, as can be seen, our bounds are very liberal and more careful analysis can help improve this bound into a more interesting form.

Another problem is that unlike Euclidean spaces, graphs can be highly non-uniform. So we restrict our attention to graphs which are uniform in terms of measure distribution. Imposing same volume growth laws at every vertex ensures this. There are graphs which do not fall in this category but locally, in well defined regions, satisfy the uniformity we seek. In this case one can analyze the regions separately and then try to deduce the properties of the whole graph from these components. Finally we have highly heterogenous graphs. Here, the geometric notion of dimension completely breaks down. Since these graphs arise in completely combinatorial settings, properties like dimension are much harder to attribute and justify. Coming back to the more measure uniform graphs, the question is \textit{how much geometric information can be conveyed through the transience class?}

The dimension of a graph and deciding degeneracy of subgraphs are obvious applications. The theory of Ricci flows on Reimannian manifolds has recently found immense favor in mathematical community. To develop the analogous theory on graphs, the Harnack's constant is potentially going to play significant role. To conclude, we bring together ideas from three different areas of mathematics. Our bounds demonstrate that the research in each of these areas has ramifications in the other two. In this sense, the importance of resolving any question in any of these areas is greatly magnified. We hope that this paper in conjunction with our earlier work motivates further investigation on discrete diffusions, in analogy with random walks and complex analysis.


\appendix
\section{Basic Potential Theory}\label{app:harmDisc}

For a proper introduction to harmonic functions on graphs, we refer the reader to the beautiful paper by Benjamini and Lovasz \cite{BL03}. See Telcs \cite{AT} for a thorough view. We start with some important definitions and fundamental properties. Given a connected graph $G$ and a function $\pi:V(G) \rightarrow \mathbb{R}$, we say that $\pi$ is {\em{harmonic}} over the vertex set $V_{h}$ if,

\begin{eqnarray}
 \frac{1}{\text{degree}(v)}\sum_{u \sim v}\pi(u) = \pi(v) \quad v \in V_{h} \nonumber
\end{eqnarray}

The vertices in $V - V_{h}$, adjacent to any vertex in $V_{h}$, are called the ``poles'' of $\pi$. The set $V_{h}$ is also called the {\em{interior}} of $\pi$ and the set of poles referred to as the {\em{boundary}}. Being harmonic over $V_{h}$ means that the value of $\pi$ at any vertex in $V_{h}$ is the average of its value in the immediate neighborhood. In case of multi graphs, we take the appropriate weighted means, where the weights are the number of common edges. This leads us to the first basic property,

\begin{prop}\label{prop:potMinMax} Any non-constant harmonic function can assume its extreme values only at the set of poles.\end{prop}

It follows that every non-constant harmonic function has at least two poles, its maxima and minima. Such functions are completely determined by their values on these vertices. Formally speaking,

\begin{prop}\label{prop:harm:uni1} Uniqueness: If two functions harmonic on $V_{h}$ agree on the boundary, they agree everywhere in the interior.\end{prop}

Property \ref{prop:harm:uni1} is important as it allows one considerable freedom in constructing harmonic completions of functions defined over the boundary set. This problem is the discrete analogue of the classical boundary value problems in complex analysis. If one fixes the interior set $V_{h}$ and allows arbitrary boundary values, the complete set of harmonic functions is obtained. This set is closed under linear combinations and contains all the constant functions.

\begin{prop}\label{prop:harm:VS} The set of functions harmonic over any set $V_{h}$ form a vector space.\end{prop}


Two important scenarios in which these functions arise naturally are electric networks and random walks.

\textit{Electric Networks:} Consider a resistive electric network (i.e. a circuit made up entirely of resistors). Let $_{s}\pi_{t}(v)$ be the potential that appears at node $v$ when unit potential is applied across $t$ and $s$. Using the equation of charge conservation (Kirchoff's node law), one can show that these potentials are harmonic on all nodes except $s$ and $t$.

\textit{Random Walks on Graphs- dipole version:} Consider a graph $G$ and two special vertices $s$ and $t$. The potential associated with $v$, with $s$ and $t$ as poles, $_{s}\pi_{t}(v)$ is defined as the probability of reaching $t$ before $s$ starting from $v$. One can check that the function $\pi$ so defined is indeed harmonic on the set $V - \{ s,t\}$, with the maximum value of $1$ at the node $t$ and the minimum value $0$ at $s$. The generalization to the multi-pole situation is simple, but it is interesting for a different reason.

\textit{Random Walks on Graphs- multipole version:} Again we consider graph $G$, but this time we have a set of interior nodes $V_{h}$ and the corresponding set of poles, $P = \{s_{i}\}$. Denote by $\pi_{j}(v)$, the probability of a random walk, starting at $v$, hitting $s_{j}$ before any other vertex in $S$. Assign any desired set of values $\pi(s_{i})$ to each of these poles. The value $\pi(v)$ is defined as the sum $\sum_{j} \pi(s_{j})\pi_{j}(v)$. As before, it is a simple exercise to check the function $\pi$ is harmonic over $V_{h}$. The formula basically suggests taking an expectation over the boundary values, using the measure induced by the random walk.

The main implication here is that one can intuitively think of the electric network theory as an analysis of random walks of electrons on the underlying graphs. Consequently, results from network theory can be used to prove interesting facts in other related areas. As an example, consider the problem of constructing the harmonic completion of a function with given boundary values. All one needs to do is to take the corresponding circuit and apply potentials equal to the boundary values on the boundary points. The potentials that will appear on other nodes can be computed using basic linear algebra (the only non-trivial step involves inverting the combinatorial Laplacian of $G$) thus allowing construction of harmonic completions efficiently.

The reciprocity theorem can be restated in terms of just potential sources and potential measurements using the notion of effective resistances between pairs of nodes. The effective resistance between a pair of nodes $u$ and $v$, $R_{eff}(u,v)$ is defined as the potential difference which develops between $u$ and $v$ if a unit current source is applied across $u$ and $v$. In any resistive network, the following reciprocity property holds. See \cite{CS11} for an elementary proof.

\begin{lem}\label{lem:potRec}Potential Reciprocity Lemma : If taking $s$ and $t$ as poles with $\pi(s) = 0$ and $\pi(t) = 1$ induces a potential of $_{s}\pi_{t}(v)$ at node $v$ and interchanging the roles of $v$ and $t$ induces $_{s}\pi_{v}(t)$ at $t$ then,
\begin{eqnarray}
 R_{eff}(s,t)_{s}\pi_{t}(v) =  R_{eff}(s,v)_{s}\pi_{v}(t)
\end{eqnarray}\end{lem}

In particular, when the effective resistances across $s$ and $t$ are the same as $s$ and $v$, we have $_{s}\pi_{t}(v)  =   _{s}\pi_{v}(t)$. In the following discussion, we will omit the left subscript ($s$) from  $_{s}\pi_{t}$ whenever it is clear from context. We say that \textit{a walk $P$ is an instance of $_{s}\pi_{t}$} if it starts at some vertex $v$, avoids $s$ and ends at $t$. The following lemma may already be known to experts. A proof appears in \cite{CS11}.

\begin{lem}\label{lem:potTri} A triangle inequality for potentials
\begin{eqnarray}
\pi_{i}(j).\pi_{j}(k) \leq \pi_{i}(k)
\end{eqnarray}
\end{lem}


\section{Harmonic functions in continuous spaces}\label{app:harmCont}

\begin{figure}[ht]
\centering
\includegraphics[width=.3\textwidth]{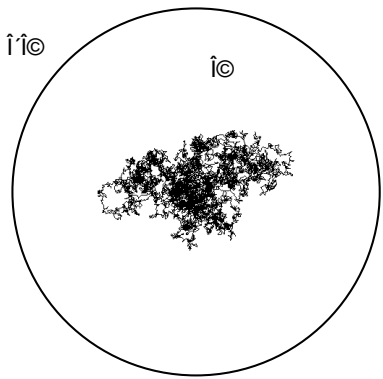}
\caption{A region $\Omega$ inside $\mathbb{R}^{2}$: Brownian motion starting at the center $p$ eventually hits the boundary $\delta \Omega$}
\label{fig:contReg}
\end{figure}

In the introductory section on potential theory on graphs, we briefly presented an example of multi-pole based construction of harmonic functions. Here we continue that discussion in the setting of real and complex analysis on continuous spaces. In analogy with the graphs induced on the vertex set $V_{h}$ and the boundary sets connecting the poles to interior, we talk about open sets in $\mathbb{R}^{n}$ and the boundary of these sets. For the sake of illustration, consider an open set $\Omega \subset \mathbb{R}^{2}$ and its boundary set $\delta \Omega$. For definiteness, let $\Omega = B(p,r)$ that is a ball of radius $r$ around the point $p$ (Figure (\ref{fig:contReg})). Consider a function $h :\delta \Omega \rightarrow \mathbb{R}^{+}$, that is an assignment of non-negative values to the boundary of this ball. Consider a real valued stochastic process $\{ (B_{1}(t),B_{2}(t)) : t \geq 0 \} $ with $B_{1}$ and $B_{2}$ independent and satisfying the following properties.

\begin{itemize}
 \item $(B_{1}(0),B_{2}(0)) = p$, the process starts at $p$.
 \item For any pair of non-overlapping time intervals $0 \leq t_{1} \leq t_{2} \leq t_{3} \leq t_{4}$, the increments, in both coordinates, $B_{i}(t_{2}) - B_{i}(t_{1})$ and $B_{i}(t_{4}) - B_{i}(t_{3})$ are independent.
 \item For all $t\geq 0$ and $h > 0$, $B_{i}(t + h) - B_{i}(t)$ is normally distributed with mean zero and variance $h$.
 \item The sample paths of this process are almost surely continuous.
\end{itemize}

Above are the defining properties of a Brownian motion in two dimensions. We refer the reader to the recent comprehensive monograph by M\"{o}rters and Peres \cite{MP10} for a thorough discussion of the properties of Brownian motions and its analogies with random walks. Just like a random walk starting at some point hits any node on the boundary with some probability, a Brownian motion starting at some point hits any subset of the boundary with some probability. Using these probabilities, we can infer the distribution and, hence, the density of the measure induced by the Brownian motion on the boundary set. Note that the density is a function of the starting point of the Brownian motion. We now define the \textit{harmonic continuation} of $h$ over the region $\Omega$. For every point $q$ inside the ball, the value $h(q)$ is defined as the expected value induced by this density on the boundary.

A function $h$ is called harmonic over $\Omega$ iff it satisfies the Laplace equation (\ref{equ:Lap}) over all points \textit{inside} $\Omega$.

\begin{eqnarray}\label{equ:Lap}
\triangledown^{2}h = \frac{\delta^{2}h}{\delta x_{1}^{2}} + \frac{\delta^{2}h}{\delta x_{2}^{2}} + \ldots + \frac{\delta^{2}h}{\delta x_{n}^{2}} = 0
\end{eqnarray}

One can also use the equivalent formulation of harmonic function in terms of its mean value property.

\noindent \textbf{Mean Value property of harmonic functions:} If $B(p,r)$ is the closed ball of radius $r$, centered at $p$, and completely inside $\Omega$ on which the harmonic function $h$ is defined, then the value at the center of the ball is given by the average value at the surface of the ball.

\begin{eqnarray}\label{equ:harmInt}
 h(x) = \frac{1}{|S^{n-1}|} \int_{\delta B(p,r)}h(q)d\sigma
\end{eqnarray}

Here $|S^{n-1}|$ is the total volume of the surface of the unit ball in $n$-dimensions and $d\sigma$ is the $n-1$ dimensional surface measure. Conversely, all integrable functions satisfying the mean value property are harmonic.

The equivalence of these conditions is discussed at length in \cite{MP10}. In the context of graphs we refer the reader to the survey by Hoory, Linial and Wigderson  \cite{LHW06}. They discuss and derive the concept of a Laplace operator in the context of graphs by first considering appropriate versions of divergence and gradient. Taking the divergence of the gradient, in the usual manner, yields the Laplacian. The linear operator corresponding to this is, not surprisingly, known as the combinatorial Laplacian of the graph. Once the formulations are in place, establishing the equivalence of mean value property and the Laplace equation is trivial.

In our discussion till now, we have overlooked a very serious issue relating to boundary values as well as the boundary structure. The formulation in terms of Laplacian is local while that in terms of mean value is dependent on boundary values. The mean value formulation depends on the integral computation, which places geometric constraints on the structure of boundary sets. The property which the boundary is required to follow is called Poincar\'{e} cone condition. (see \cite{MP10}, Definition 3.10). Once we ensure that the boundary set is well behaved, we also need a guarantee that for given boundary values, a harmonic continuation will exist. Indeed, this is the classical Dirichlet boundary value problem. Fortunately, the only condition required is that the values are continuous. The uniqueness of solution is easy to show. In fact the set of harmonic functions over a domain form a vector space. For a full discussion we refer the reader to the classical text by Rudin \cite{WR}. The most important examples of harmonic functions come from the real and imaginary parts of holomorphic (i.e. complex differentiable) functions. In this case, harmonicity of these functions is a simple consequence of the Cauchy-Riemann equations which every holomorphic function must satisfy.

Following above discussion, it is reasonable to expect that a harmonic function should satisfy some non-trivial global properties. One such property is elucidated by the celebrated Harnack's inequality. We present below the two dimensional version for balls.

\textit{Harnack's inequality for balls:} If $h$ is positive harmonic function defined over $\Omega$ then the following inequality holds for any (open) ball $B(p,r) \subseteq \Omega$ and $q \in B(p,r)$.

\begin{eqnarray}\label{equ:harnBall}
 h(q) \leq \left( \frac{r}{r - |p - q|} \right)^{2}h(p)
\end{eqnarray}

The generalization to higher dimensions is straightforward. From this, the inequality for all compact connected regions $K ( \subseteq \Omega)$ follows using a very simple chaining argument which involves considering a set of open balls in $K$ which cover the path from the reference vertex $p$ to the vertex $q$. Using the bounds from equation (\ref{equ:harnBall}), the growth rate of harmonic functions in each of these balls is bounded by the respective constants $H_{i}$. Multiplying these gives us an obvious bound over the considered path. By taking any finite covering of $K$ (compactness of $K$ guarantees the existence of such a covering), one can find an absolute bound on the value of the Harnack's constant by simply multiplying the constants corresponding to each of the balls (see figure (\ref{fig:harnCov-main})). This implies the Harnack's inequality over any compact region.

\textit{Harnack's inequality over general regions:} Suppose that $\Omega$ is connected and $K$ is a compact subset. Then for any harmonic function, $h$, defined over $\Omega$, the following inequality holds

\begin{eqnarray}\label{equ:harnComp}
 h(q) \leq H_{K} h(p)
\end{eqnarray}

Here $H_{K}$ is the Harnack's constant of the region $K$ and is independent of the function $h$. Note that the version for balls (\ref{equ:harnBall}) uses the exact value for this constant. This value comes from an alternate formulation of the integral equation (\ref{equ:harmInt}) based on the Poisson kernel. For further discussion on the Poisson kernel, we refer the reader to Rudin \cite{WR}. Harnack's inequality plays fundamental role in the traditional complex analysis and PDE theory. Recently it found use in Perelman's proof of the Poincar\'{e} conjecture.

\end{document}